# Applying Systems-Theoretic Process Analysis in the Context of Cooperative Driving

Joakim Oscarsson, Max Stolz-Sundnes, Naveen Mohan, Viacheslav Izosimov
KTH Royal Institute of Technology
Stockholm, Sweden
Corresponding Author email: naveenm@kth.se

*Abstract*— Highly automated, cooperative driving vehicles will allow for a more fluid flow of traffic, resulting in more efficient, eco-friendly and safe traffic situations. The automotive industry however, is safety critical and current safety standards were not designed to deal with cooperative driving. In this paper, we apply a modern safety analysis method, Systems-Theoretic Process Analysis, in the context of cooperative driving as part of the Grand Cooperative Driving Challenge (GCDC) and present our reflections on the method.

## I. INTRODUCTION

The second edition of the Grand Cooperative Driving Challenge (GCDC) will be held in May 2016. It is a competition aiming to speed up real-life implementation of cooperative driving. Cooperative driving will optimize traffic flow, reduce emissions and increase safety. KTH Royal Institute of Technology is competing in the GCDC with two vehicles: A Research Concept Vehicle (RCV) and a Scania truck. Since the automation of propulsion functionality is highly safety critical, a thorough safety analysis is required. In this case, the analysis has been performed using Systems-Theoretic Process Analysis (STPA).

A Cooperative Driving Module, responsible for control of the vehicle and for communicating with surrounding vehicles, is currently being designed at KTH. This will be instantiated in both the vehicles individually for implementation.

To the best of our knowledge, this is the first paper on STPA in a cooperative driving context. As such, the unique contributions of this paper are *(1) analysing the benefits of applying the STPA method in this context* and *(2) sharing of the experiences from our work.*

The remainder of this paper is structured as follows. Section II sets the background and describes the GCDC, the RCV, automotive safety and the STPA method. Section III describes the given task and the process followed. Section IV presents the preliminary findings, and section V discusses the process and gives the authors' reflections of the STPA process as well as a description of future work.

## II. BACKGROUND & RELATED WORK

### A. Grand Cooperative Driving Challenge

Ten university teams are scheduled to compete in the upcoming edition of GCDC, organized by the i-GAME research project [1]. For all vehicles, cooperation is to be achieved through vehicle-to-vehicle and vehicle-to-infrastructure communication, from here on referred to collectively as V2X, which utilizes the work being done in Cooperative Intelligent Transport Systems (C-ITS), facilitated by the European Telecommunications Standards Institute (ETSI). The competition includes three scenarios. Two for the actual competition and one for demonstration purposes. All scenarios are to be performed with consideration of comfort and safety of passengers in mind without unnecessary loss of speed. The scenarios include (1) the merging of two convoys, (2) a T-intersection and (3) yielding to prioritised vehicles, such as an ambulance, as the demonstration scenario [2].

### B. Research Concept Vehicle

The RCV, shown in Fig. 1 is a drive-by-wire electric vehicle developed at the Integrated Transport Research Lab (ITRL) at KTH with the purpose of validating and demonstrating current and future research. It is based on the concept of "autonomous corner modules", i.e. wheels with built-in electric hub motors and actuators for active camber tilt and yaw control of each wheel. The addition of a battery and a central computing unit to control the modules, results in a fully electric drive train. The vehicle, in its current design, can reach a top speed of approximately 55 km/h on a flat surface. In order to comply with the GCDC communication standards, the RCV will be fitted with a Vehicle Intelligent Transport Systems Station.

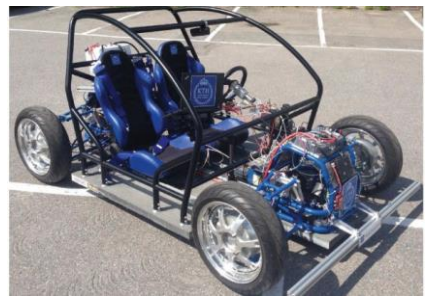

Fig. 1. Research Concept Vehicle

## C. Automotive Safety

The best practices in automotive functional safety today recommend complying with the ISO-26262 standard [3]. This standard is currently applicable to vehicles that include at least one electrical and/or electronic (E/E) system in series produced passenger vehicles up weighing up to 3500 kg. The standard concerns the functional safety of E/E systems. Regarding the concept phase of the lifecycle, ISO-26262 requires: A hazard analysis, a risk assessment and an Automotive Safety Integrity Level (ASIL) determination [4], the purpose of which is to develop a safe vehicle. Following the ISO-26262 safety lifecycle, a safety analysis is to be conducted. The objective of which is to investigate the consequences of faults and failures, and to identify new hazards that were not detected during the hazard analysis [5]. Some examples of methods advocated by the standard are Hazards and Operability analysis (HAZOP) [6], Failure Modes and Effects Analysis (FMEA) and Fault Tree Analysis (FTA) are explained in greater detail in [7]. However, the concept of cooperative driving is relatively new, and neither the ISO-26262 nor these techniques, were designed with a scope of multiple vehicles in mind. Designing systems for cooperative driving poses different challenges than singular vehicles from a safety viewpoint.

## D. STPA

Systems-Theoretic Process Analysis (STPA) is a safety analysis method developed by Nancy Leveson as a response to the raising complexity and software dependencies of modern products. In her book, Engineering a Safer World [8], she argues that traditional causality models are based on chains of failures, and that traditional safety analysis methods try to ensure safety by preventing these failures. This is, according to her, becoming an outdated way of thinking about safety since modern products are so complex that undetected design flaws are a real safety issue. Components may function as intended, and nothing may fail, but the system could still be unsafe due to unforeseen behaviours caused by the complex interactions between components and subsystems. This scenario is of particular importance to consider in the context of cooperative driving since all vehicles, need not necessarily even be designed by the same company, while their interactions are essential to provide collaborative functioning. STPA is based on the causality model Systems-Theoretic Accident Model and Processes (STAMP), which is in turn based on systems engineering. Before the STPA process can begin, a STAMP model must be created. This includes first defining accidents to avoid and hazards that could cause them. Then, the hazards are translated into high-level safety constraints (SC) that can be used as aid in the making of a preliminary architecture. To complete the STAMP model, the functionality then needs to be represented as a control structure. A control structure consists of one or more control loops consisting in turn of a controller, which is controlling some process through actuators, getting feedback through sensors. The means by which the controller is controlling the process are called control actions. A generic control loop is illustrated in Fig. 2.

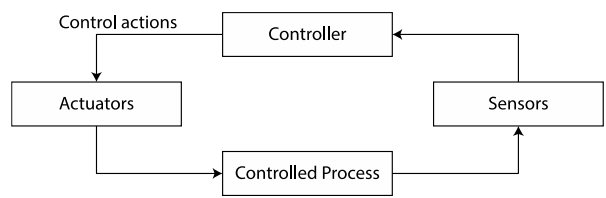

Fig. 2. Simplified standard control loop

The process of STPA consists of two steps. The first step corresponds to a hazard analysis. Its purpose is to determine when a specific control action can lead to a hazard and thereby to a defined accident. The control actions of each control loop are checked against four different ways in which the action could potentially lead to a hazard. If a scenario exists in which a control action could lead to a hazard, then that control action is considered unsafe. The four ways an unsafe control action (UCA) could be termed unsafe are:

a) If providing a control action would lead to a hazard.
b) If not providing a control action would lead to a hazard.
c) If providing a control action too early / too late / out of sequence would lead to a hazard.
d) If providing a control action for too long / stopped too soon would lead to a hazard.

UCAs, when identified, are reformulated into SCs. The second step of STPA corresponds to a causal analysis. The purpose of this step is to find all *potential causes* of the UCAs derived from STPA step 1. The reason why it is not referred to as a fault analysis is to emphasise the fact that failures are not the only potential causes. Hence STPA step 2 is more than just a fault analysis. When causes are identified, measures to mitigate them are to be taken. If such measures are not possible, then STPA step 2 can be iterated until the causes are broken down sufficiently, such that they can be dealt with. Fig. 3. visualises the STAMP-STPA process.

Researchers have previously studied STPA in the automotive context. For example, Kawabe and Yanagisawa applied STPA on the human interaction with a four-wheel drive power-train, and found that it was a much faster process compared to the other alternatives [9]. Abdulkhaleq and Wagner applied STPA to a cruise control prototype, linking together STPA code level safety verification [10].

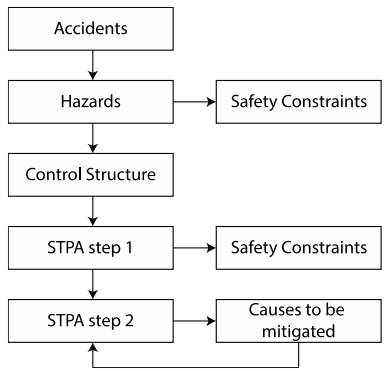

Fig. 3. Overview of the STPA process

Other related work include Asare et al. who formalised a more rigorous framework for STPA part 1, called FSTPA-I, to make the STPA process less ambiguous [11].

### III. TASK AND STPA PROCESS DESCRIPTION

The task to conduct a safety analysis of the Cooperative Driving Module is a joint task of two different studies. The first studies how STPA could be adapted into the ISO-26262 lifecycle, and comparing the impact of safety analysis on the architecture of the vehicle, on the basis of the scope and the context of the analysis (singular vehicle context versus a cooperative multi-vehicle context).

The second study takes a closer look at the safety aspects involving the driver. The work aims to answer how system modes, driving mode switching and safety related HMI aspects can be identified and quantified using the results obtained with the STPA method in the first task as well as how these aspects of the cooperative functionality should be presented to the driver.

Both projects use the same real world cases for validation; the development of the RCV and the Cooperative Driving Module. It implements communication with other vehicles though V2X, as specified by the C-ITS standard [12], with some extensions developed by the i-GAME project [13]. Based on the input from vehicle sensors and the cooperative effort of the surrounding vehicles, the purpose is longitudinal and lateral control over the vehicle. Since these are safety critical functionalities, a safety analysis covering the cooperative driving module is required for participation in the GCDC.

| Accidents | | Hazards | |
|---|---|---|---|
| A1 | Collision with vehicle | H1 | Inadequate distance to frontal vehicle |
| | | H2 | Inadequate distance to rear vehicle |
| | | H3 | Inadequate distance to side vehicle |
| A2 | Collision with environment | H4 | Inadequate distance to frontal environmental object |
| | | H5 | Inadequate distance to side environmental object |
| A3 | Driver G-force too high | H6 | Acceleration too strong |
| | | H7 | Deceleration to strong |

The safety analysis method of choice, STPA, was chosen because it is a relatively new method (2011) that has not been studied in great detail in the automotive context, and it was developed to match the complexity of modern software intensive products [8], unlike the other alternatives mentioned in the background section which were designed in the 1950s and the 1960s. Even though STPA covers a broad view of system safety including environmental and organisational changes, the scope was limited to functional changes only for this study.

The analysis was led by the two primary authors of this paper, and conducted in collaboration with the RCV development team at ITRL and two safety experts (who are co-authoring this paper) from the mechatronics division at KTH. Workshops were held with the development team regarding the accidents and hazards definitions, the STPA Step 1 and the STPA Step 2, with intermediate analysis sessions by the lead analysts. During these sessions the project leader and lead architect provided continuous feedback and verification of the analysis. The number of workshop attendants was between four and six, depending on the specific workshop.

### IV. FINDINGS

To initialize the STPA process, accidents were defined for the RCV's participation in the GCDC scenarios. Each accident was then broken down into hazards, resulting in seven primary hazards. The accidents and hazards considered can be seen in Table 1.

TABLE I.    DEFINITIONS OF ACCIDENTS AND HAZARDS

Each hazard was then translated into corresponding high level SCs for the system. After iterative development together with the team developing the extended functionality of the RCV, an architecture model of the cooperative module was created which defines the system and the functionalities to be analyzed. At the top level, the system architecture consists of:

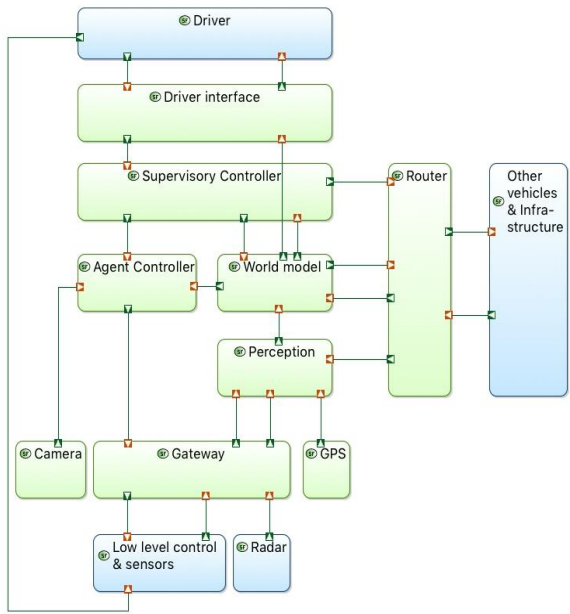

- A perception functionality responsible for estimating the state of the vehicle and of the surrounding vehicles.

- A 'world model' which acts as an onboard logger and database, containing state estimates.

- A router to route V2X transmissions.

- A supervisory control functionality, which has the main responsibility of high level decision making and control.

- An agent controller, hosting various control agents, which controls the platform (ego-vehicle).

- A gateway to separate the communication internal to the cooperative driving module, from the rest of the ego-vehicle.

- An HMI for driver input / output.

The supervisory controller controls the system by configuring the agent controller, and activates the appropriate control agents, depending on the situation. These control agents in turn send reference values to the low level drivers of the Ego-vehicle. Sensor data from the platform is fused with incoming communication from other vehicles in the perception layer to create an estimate of the state of the vehicle and its surroundings. All estimates and received transmissions are being stored in the world model and can be accessed when needed. The cooperative module is delimited by: the gateway towards the own vehicle, the router towards other vehicles and the HMI towards the driver. The modelled architecture can be seen in Fig. 4.

All relevant controllers and corresponding control loops in the system were identified. Nine control loops were defined: driver to supervisory controller, and eight loops related to sub

Fig. 4. System architecture of the cooperative module

functions of the agent control functionality and the supervisory controller or low level control.

STPA Step 1 was performed in collaboration with domain experts during two workshops, resulting in 98 UCAs. For example, one detected UCA was: "Supervisory controller not providing a reference vehicle when activating the cooperative adaptive cruise control agent". That is, the adaptive cruise control would not receive a target vehicle to adapt the speed to, which could lead to hazards H1, H2, and H3 (See Table 1)

The remaining UCAs were translated into SCs, which were considered equivalent to ISO-26262's safety goals, and assigned ASILs assessing: Severity (S0-S3), exposure (E0-E4) and controllability (C0-C3) to match the ISO-26262 process. Due to the relative top speed of 60 km/h in scenario two and the lack of personal safety features such as airbags on the RCV, a majority of the SCs were given the highest severity level S3, meaning possibly life threatening injuries. Exposure level was set at E4 for most constraints seeing that practically all analyzed controllers are used in performing every scenario in the competition, resulting in a high probability of exposure. All considered SCs were given a controllability level of C3 since the system is designed to ideally perform the scenarios without driver interaction and no backup systems currently exist. As a result, the SCs were with few exceptions all assigned ASIL D, which is the integrity level that corresponds to the highest harm [4].

The findings in step 1 were put through step 2 of the STPA process in collaboration with experts analyzing each UCA and looking at the related control loop to reveal any causal factors and/or specific scenarios which could result in unsafe control of the vehicle. One example of an identified causal scenario for the UCA mentioned in the pervious example is if the both the V2V communication and the perception functionality would fail at the same time causing the own estimate of surrounding vehicles to be wrong at the same time as their communication not reaching through.

All resulting SCs related to specific control loops, controllers and control actions were presented to the RCV development team to be taken into consideration during further development together with a list of recommendations for future work concluded from the results of the causal analysis. These recommendations highlighted the parts and issues of the system in which flaws or errors would often appear in the second step of the analysis and could be involved in contributing to a large number of UCAs. For example; the system in its current state is in need of some way of detecting loss of traction.

## V. DISCUSSION AND CONCLUSIONS

The results from the STPA analysis has been submitted to the GCDC organizers for a review before the competition. We will in this section discuss the results and give our own reflections of the process that we conducted.

The high level SCs derived from the system level hazards were in this case not used to form a basis for the concept architecture since a preliminary, although not complete, architecture already existed.

The preliminary system level architecture changed several times during the control structure creation process. As parts of the control structure were verified flaws were detected and the architecture had to be reviewed. Due to the agile work style of the development team, these flaws and many identified UCAs could be addressed on the fly through architectural changes. It is our opinion that the STPA process quickly refined and improved the architectural design where such design already existed, and strongly contributed to the parts that were undefined at the start of the analysis. Such as the data flow between the world model and some control agents, and the responsibilities of the supervisory controller respectively.

The part that we perceived as the most difficult one was the STPA step 2. It was the most time consuming and the most repetitive part. The rest of the STAMP-STPA method was well structured, easy to follow and gave a clear picture of the completeness of the analysis. For step 2 however, even though example lists of commonly occurring causes were at the team's disposal, the process would best be described as a brain storming session which ended when no one could come up with any more causes. Because of this, the completeness depends on a combination of expertise and imagination of the participants, which is the same critique other safety analysis methods have received. Thus, it would seem like more research on structuring STPA step 2 is needed.

In general, in spite of the criticism discussed above, we found that STPA was well suited to our purposes and we could efficiently utilize it for the problem of safety analysis of the cooperative driving functionality. Thus, we see a good potential for STPA in other complex systems.

ACKNOWLEDGEMENT

Support from FFI, Vehicle Strategic Research and Innovation and Vinnova through the ARCHER (proj. No. 2014-06260) and FUSE (proj. No. 2013-02650) projects, is acknowledged.